\documentclass{main}

\usepackage{amsmath}
\usepackage{amssymb}

\newcommand{\Photo}{}

\begin{document}

\title{\Large Collision geometry in UPC dijet production}

\author{Kari J.\ Eskola, Vadim Guzey, Ilkka Helenius, Petja Paakkinen\,\footnote{Speaker, email: petja.k.m.paakkinen@jyu.fi} and Hannu Paukkunen}

\address{University of Jyväskylä, Department of Physics, P.O. Box 35, FI-40014 University of Jyväskylä, Finland\\ and Helsinki Institute of Physics, P.O. Box 64, FI-00014 University of Helsinki, Finland}

\maketitle\abstracts{
We present a comprehensive NLO pQCD study on inclusive dijet photoproduction in ultraperipheral nucleus-nucleus collisions (UPCs). Our analysis takes into account the finite size of both the photon-emitting and the target nucleus, estimated using the Wood-Saxon nuclear density profile. We show that a significant part of the measured dijets at large $z_\gamma$ in UPC Pb+Pb collisions at 5.02 TeV come from events with relatively small impact parameters of the order of a few nuclear radii, and the cross section predictions thus become sensitive to the modelling of the collision geometry and photon flux near the source nucleus. In addition, we include the modelling of electromagnetic breakup survival factor needed for a direct comparison with the experimental data and study the resolution power of this process in terms of the impact-parameter dependent nuclear parton distribution functions.
}


\section{Introduction}

In ultraperipheral nucleus-nucleus collisions (UPCs), inclusive dijet photoproduction has been proposed as a valuable probe for studying nuclear parton distribution functions (nPDFs).\cite{Strikman:2005yv,Guzey:2018dlm,Guzey:2019kik} Compared to e.g.\ jet production in proton-nucleus collisions this offers an arguably cleaner probe with considerably smaller underlying event activity. However, the unique condition of UPCs---specifically, the absence of nuclear overlap---imposes restrictions on the impact parameter space. This limitation becomes particularly relevant in dijet production, where the requirement for high-transverse-momentum jets necessitates an energetic photon in the initial state. Such photons are more likely to originate from close to the source nucleus, and, consequently, the cross section predictions for dijet production become sensitive to the modeling of nuclear geometry and the photon flux near the source nucleus.

Here, we present a study on the inclusive UPC dijet photoproduction using next-to-leading order (NLO) perturbative QCD (pQCD) and the impact-parameter dependent equivalent photon approximation (EPA).\cite{Baron:1993nk,Greiner:1994db,Krauss:1997vr} The study considers the finite size of both the photon-emitting and the target nucleus, revealing a sensitivity to the transverse-plane geometry of the UPC events.\cite{Eskola:2024fhf} We discuss also the role of the forward-neutron event-class selection in isolating the photonuclear cross section and include the associated probability for no electromagnetic (e.m.) breakup of the photon-emitting nucleus in the predictions. We show that the geometrical effects survive even after including this additional suppression factor. Full details of this work can be found in Ref.~\citelow{Eskola:2024fhf}.

\section{Dijet production in impact-parameter dependent EPA}

We work here in terms of the impact-parameter dependent EPA,\cite{Baron:1993nk,Greiner:1994db,Krauss:1997vr} where the UPC dijet cross section may be written as
\begin{multline}
    {\rm d} \sigma^{AB \rightarrow A + {\rm dijet} + X} = \sum_{i,j,X'} \int {\rm d}^2{\bf b} \, \Gamma_{AB}({\bf b}) \int {\rm d}^2{\bf r} \, f_{\gamma/A}(y,{\bf r}) \otimes f_{i/\gamma}(x_\gamma,Q^2) \otimes \int {\rm d}^2{\bf s} \, f_{j/B}(x,Q^2,{\bf s}) \\ \otimes {\rm d} \hat{\sigma}^{ij \rightarrow {\rm dijet} + X'}(x_\gamma y p_A,x p_B,Q^2) \times \delta^{(2)}({\bf r} - {\bf s} - {\bf b}).
  \label{eq:xsec_full}
\end{multline}
Here, $f_{\gamma/A}(y,{\bf r})$ is the flux of photons from a source $A$ carrying a fraction $y$ of the per nucleon beam momentum $p_A$, evaluated at a transverse distance $|{\bf r}|$ from the center of $A$. A photon represented by this flux interacts with a parton $j$ from the target $B$ at a distance $|{\bf s}|$ from its center, with the associated impact-parameter dependent nPDF denoted as $f_{j/B}(x,Q^2,{\bf s})$ for a parton with a fraction $x$ of the per nucleon beam momentum $p_B$. This interaction, characterised by the partonic cross section $\hat{\sigma}^{ij \rightarrow {\rm dijet} + X'}(x_\gamma y p_A,x p_B,Q^2)$ with a hard scale $Q^2$, can happen either directly, where $i = \gamma$ and $f_{\gamma/\gamma}(x_\gamma,Q^2) = \delta(1 - x_\gamma)$, or through a resolved contribution, with $i = q,\bar{q},g$ and $f_{i/\gamma}(x_\gamma,Q^2)$ being the corresponding photon parton distribution for a parton carrying a fraction $x_\gamma$ of the photon energy. Finally, the survival probability $\Gamma_{AB}({\bf b})$ requires that there is no hadronic interaction between the nuclei $A$ and $B$ at the impact parameter ${\bf b} = {\bf r} - {\bf s}$.

\begin{figure}
  \raisebox{-0.5\height}{\includegraphics[width=0.495\textwidth]{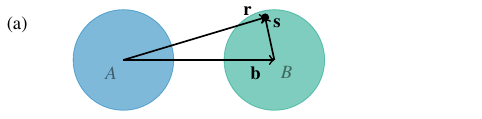}}
  \raisebox{-0.5\height}{\includegraphics[width=0.495\textwidth]{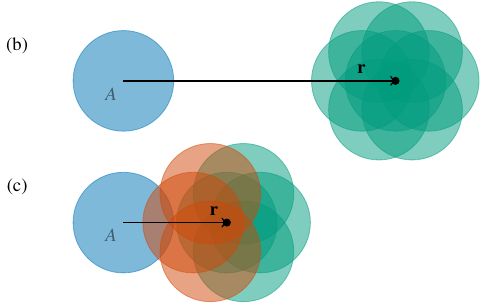}}
  \caption{(a) An illustration of the transverse-plane vectors ${\bf r}$, ${\bf b}$ and ${\bf s}$. (b) A `far-passing', large $|{\bf r}|$ scenario. (c) A `near-encounter', small $|{\bf r}|$ scenario. Figure from Ref.~\protect\citelow{Eskola:2024fhf}.}
  \label{fig:transverse_plane}
\end{figure}

The collision geometry is illustrated in Fig.~\ref{fig:transverse_plane}, panel (a). The phase space can be split into two regions, cf.\ the discussion in the following section: For large $|{\bf r}| \sim |{\bf b}| \gg |{\bf s}|$, Fig.~\ref{fig:transverse_plane}, panel (b), the two nuclei never overlap, and thus any value of $|{\bf s}| < R_B$ is allowed in this `far-passing' region and the spatial dependence effectively integrates out. For $|{\bf r}| \sim |{\bf b}| \sim |{\bf s}|$ instead, Fig.~\ref{fig:transverse_plane}, panel (c), some configurations lead to nuclear overlap and are thus excluded from the UPC cross section. As we will show, this restriction in the transverse phase space for these `near-encounter' events leads to a sensitivity on the transverse-space geometry of the colliding nuclei in the UPC dijet cross section.

\section{Effective photon flux}

Using a simple assumption that the transverse and longitudinal degrees of freedom factorize for the impact-parameter dependent nPDFs, $f_{j/B}(x,Q^2,{\bf s}) = \frac{1}{B} \, T_{B}({\bf s}) \times f_{j/B}(x,Q^2)$, where $T_{B}({\bf s})$ is the nuclear thickness function, $B = \int {\rm d}^2{\bf s} \, T_{B}({\bf s})$ is the number of nucleons in the target nucleus and $f_{j/B}(x,Q^2)$ the ordinary (spatially averaged) nPDF, one can reorganise Eq.~\eqref{eq:xsec_full} to the form
\begin{equation}
  {\rm d} \sigma^{AB \rightarrow A + {\rm dijet} + X} = \sum_{i,j,X'} f_{\gamma/A}^{\rm eff}(y) \otimes f_{i/\gamma}(x_\gamma,Q^2) \otimes f_{j/B}(x,Q^2) \otimes {\rm d} \hat{\sigma}^{ij \rightarrow {\rm dijet} + X'}(x_\gamma y p_A,x p_B,Q^2)
  \label{eq:xsec_w_eff_flux}
\end{equation}
where
\begin{equation}
  f_{\gamma/A}^{\rm eff}(y) = \frac{1}{B} \int {\rm d}^2{\bf r} \int {\rm d}^2{\bf s} \, f_{\gamma/A}(y,{\bf r}) \, T_{B}({\bf s}) \, \Gamma_{AB}({\bf r}\!-\!{\bf s})
  \label{eq:eff_flux}
\end{equation}
is an effective photon flux encoding all the spatial dependence. Various approximations can be used for calculating this effective flux. We consider here the following options in order to demonstrate the importance of the geometrical effects:
\begin{description}
  \item[PL] refers to the pointlike approximation, where the finite size of the colliding nuclei is accounted for only in requiring the impact parameter to be larger than twice the nuclear hard-sphere radius, $R_{\rm PL} = 7.1\ {\rm fm}$ for a lead nucleus of $A=208$, i.e.\ $\Gamma_{AB}^{\rm PL}({\bf b}) = \theta(|{\bf b}| - 2R_{\rm PL})$. Other than that, the nuclei are treated as pointlike objects with
  \begin{equation}
    f_{\gamma/A}^{\rm PL}(y,{\bf r}) = \frac{Z^2 \alpha_{\rm e.m.}}{\pi^2} m_p^2 y [ K_1^2(\xi) + \frac{1}{\gamma_L} K_0^2(\xi) ], \quad \xi = y m_p |{\bf r}|,
    \label{eq:pl-bare-flux}
  \end{equation}
  as the bare photon flux,\cite{Bertulani:1987tz} where $Z$ is the nuclear charge, $\alpha_{\rm e.m.}$ the fine-structure constant, $m_p$ the proton mass, $\gamma_L$ the nucleus Lorentz factor and $K_{0,1}$ are modified Bessel functions of the second kind, and the target parton spatial distribution is taken as  $T_B^{\rm PL}({\bf s}) = B\delta^{(2)}({\bf s})$.
  \item[WS$_{\delta({\bf s})}$] refers to the approximation where the survival factor is obtained through the optical Glauber-model $\Gamma_{AB}^{\rm WS}({\bf b}) = \exp[-\sigma_{\rm NN} T_{AB}^{\rm WS}({\bf b})]$ with the nuclear overlap function $T_{AB}^{\rm WS}$ calculated from the Woods-Saxon distribution and $\sigma_{\rm NN}$ taken as the total (elastic+inelastic) nucleon-nucleon cross section. Likewise, the bare photon flux is taken as the one of an extended charge distribution\,\cite{Krauss:1997vr}
  \begin{equation}
    f_{\gamma/A}^{\rm WS}(y,{\bf r}) = \frac{Z^2 \alpha_{\rm e.m.}}{\pi^2} \frac{1}{y} \left| \int_0^\infty \frac{{\rm d}k_\perp k_\perp^2}{k_\perp^2 + (y m_p)^2} F_A^{\rm WS}(k_\perp^2 + (y m_p)^2) J_1(|{\bf r}|k_\perp) \right|^2,
    \label{eq:ws-bare-flux}
  \end{equation}
  where the form factor $F_A^{\rm WS}$ is again obtained from the Woods-Saxon distribution and $J_1$ is the cylindrical modified Bessel function of the first kind. For this intermediate result, we still keep $T_B^{\rm PL}({\bf s}) = B\delta^{(2)}({\bf s})$, thus neglecting the spatial distribution of partons in the target nucleus.
  \item[WS] is finally the full Woods-Saxon approximation, where in addition to the survival factor and the bare flux as in the previous case, also the nuclear thickness function for the target parton distribution is calculated from the Woods-Saxon distribution, $T_B^{\rm WS}({\bf s}) = \int_{-\infty}^\infty {\rm d}z \; \rho_B^{\rm WS}\left(\sqrt{z^2+{\bf s}^2}\right)$.
\end{description}

\begin{figure}
  \centering
  \includegraphics[width=\textwidth]{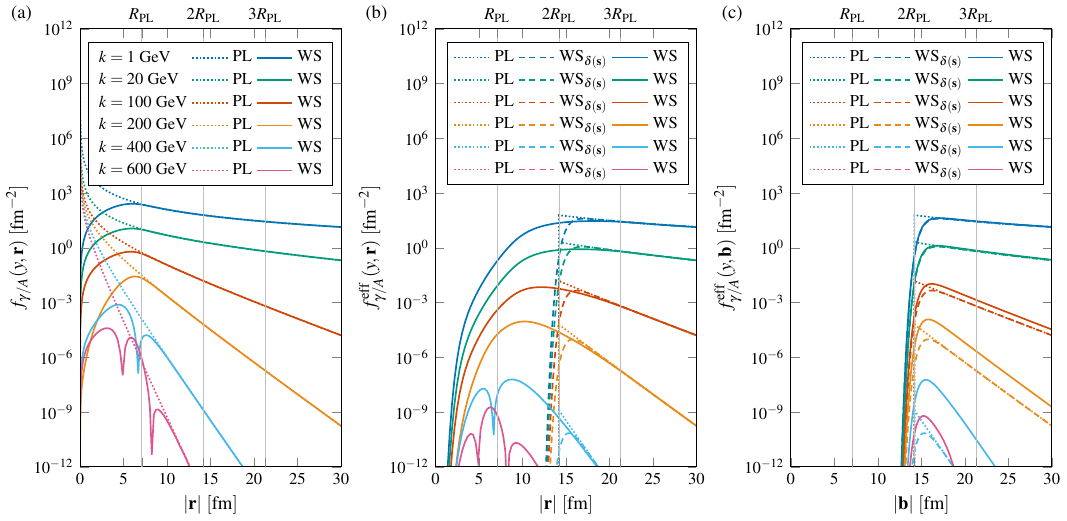}
  \caption{(a) A comparison of the bare flux from the pointlike, Eq.~\eqref{eq:pl-bare-flux}, and Woods-Saxon, Eq.~\eqref{eq:ws-bare-flux}, sources. (b) The effective $|{\bf r}|$-dependent flux calculated through Eq.~\eqref{eq:eff_flux_r_dep}. (c) The effective $|{\bf b}|$-dependent flux calculated through Eq.~\eqref{eq:eff_flux_b_dep}. Figure from Ref.~\protect\citelow{Eskola:2024fhf}.}
  \label{fig:ph-flux}
\end{figure}

In Fig.~\ref{fig:ph-flux}, panel (a), we compare the bare fluxes from the pointlike, Eq.~\eqref{eq:pl-bare-flux}, and Woods-Saxon, Eq.~\eqref{eq:ws-bare-flux}, sources. As expected, differences appear only for small values of $|{\bf r}|$. Panels (b) and (c) then show the effective flux as a function of $|{\bf r}|$ and $|{\bf b}|$, respectively, with
\begin{align}
  f_{\gamma/A}^{\rm eff}(y, {\bf r}) &= f_{\gamma/A}(y,{\bf r}) \times \frac{1}{B} \int {\rm d}^2{\bf s} \, T_{B}({\bf s}) \, \Gamma_{AB}({\bf r}\!-\!{\bf s})
  \label{eq:eff_flux_r_dep} \\
  f_{\gamma/A}^{\rm eff}(y, {\bf b}) &= \Gamma_{AB}({\bf b}) \times \frac{1}{B} \int {\rm d}^2{\bf s} \, f_{\gamma/A}(y,{\bf b}\!+\!{\bf s}) \, T_{B}({\bf s})
  \label{eq:eff_flux_b_dep}
\end{align}
satisfying $f_{\gamma/A}^{\rm eff}(y) = \int {\rm d}^2{\bf r} \, f_{\gamma/A}^{\rm eff}(y, {\bf r}) = \int {\rm d}^2{\bf b} \, f_{\gamma/A}^{\rm eff}(y, {\bf b})$. For the $|{\bf r}|$-dependent effective flux, we see that when $|{\bf r}| > 3R_{\rm PL}$, i.e.\ for `far-passing' nuclei, the three approximations described above yield an identical flux. That is, viewed from afar, any source appears as pointlike, and since in this region the probability for hadronic interaction between the nuclei is practically zero (i.e.\ $\Gamma_{AB} \approx 1$), the dependence on the target spatial distribution also integrates out in Eq.~\eqref{eq:eff_flux_r_dep}. For $|{\bf r}| < 3R_{\rm PL}$, i.e.\ `near-encounter' events, the three approximations differ due to a finite probability for nuclear overlap, and the full WS effective flux has a non-negligible contribution in the region $|{\bf r}| < 2R_{\rm PL}$ due to the integration over the full target width that is not taken into account by the other approximations. This becomes particularly important for high-energy photons, where the flux drops very fast as a function of $|{\bf r}|$. Note that if considered as a function of $|{\bf b}|$, Eq.~\eqref{eq:eff_flux_b_dep}, the integration over the target width causes always an enhancement in the WS flux over the WS$_{\delta({\bf s})}$ approximation and one cannot easily find a limit where the different fluxes would resolve to the PL approximation. See Ref.~\citelow{Eskola:2024fhf} for more details.

\section{UPC dijets in Pb+Pb at 5.02 TeV}

\begin{figure}
  \centering
  \includegraphics[width=0.495\textwidth]{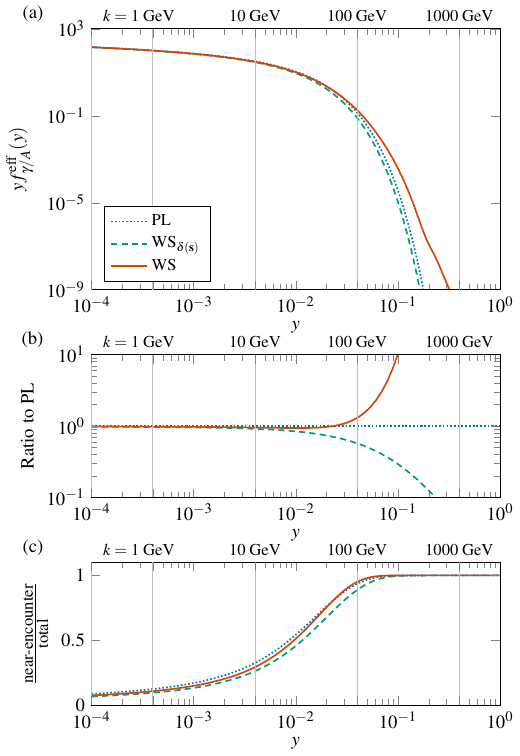}
  \includegraphics[width=0.495\textwidth]{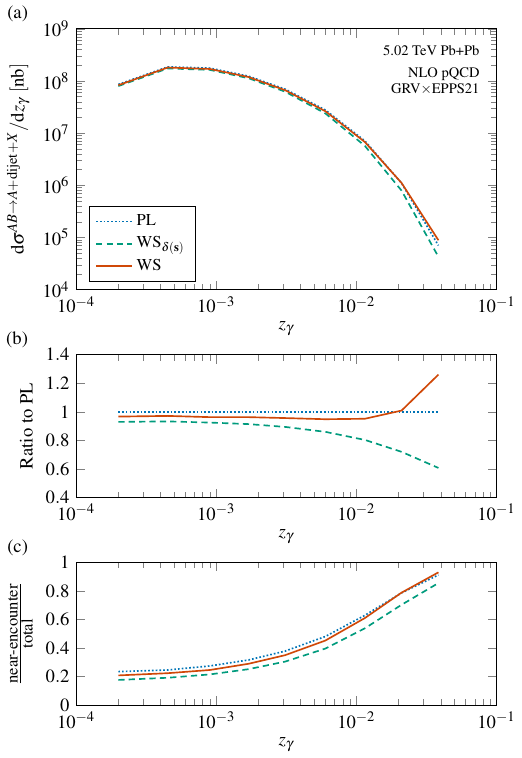}
  \caption{Comparison of the effective photon flux (left) and UPC dijet cross section (right) in Pb+Pb collisions at $\sqrt{s_{\rm NN}} = 5.02\ {\rm TeV}$. (a) Absolute quantities. (b) Ratios of the different flux approximations with respect to the pointlike one. (c) Fraction of $|{\bf r}| < 3R_{\rm PL}$ `near-encounter' versus total number of events. Figures from Ref.~\protect\citelow{Eskola:2024fhf}.}
  \label{fig:flux-vs-xsec}
\end{figure}

We compare the effective flux $f_{\gamma/A}^{\rm eff}(y)$ and the inclusive UPC dijet cross section for Pb+Pb collisions at $\sqrt{s_{\rm NN}} = 5.02\ {\rm TeV}$ in Fig.~\ref{fig:flux-vs-xsec}. For the latter, we show the single-differential distribution as a function of $z_\gamma = M_{\rm jets} \exp(y_{\rm jets}) / \sqrt{s_{\rm NN}}$, where $M_{\rm jets}$, $y_{\rm jets}$ are the invariant mass and rapidity of the anti-$k_{\rm T}$ ($R = 0.4$) jets passing the kinematical cuts. For these cuts we use the ones from the ATLAS analysis,\cite{ATLAS:2022cbd} where the jets are required to be confined in rapidity to $|\eta_{\rm jet}| < 4.4$ and have a transverse momentum of at least $p_{\rm T,jet} > 15\ {\rm GeV}$ with $M_{\rm jets} > 35\ {\rm GeV}$. At leading order, one would have $z_\gamma = x_\gamma y$ and thus this observable can be used as an experimental proxy for the cross section dependence on the photon flux. Here, we perform the calculations in NLO with the Frixione \& Ridolfi jet photoproduction code.\cite{Frixione:1997ks}

As can be seen from the figure, the effective photon flux as a function of $y$ (Fig.~\ref{fig:flux-vs-xsec}, left) and the UPC dijet cross section as a function of $z_\gamma$ (Fig.~\ref{fig:flux-vs-xsec}, right) exhibit rather common behaviour due to their intertwined nature. For low photon energies (i.e.\ small $y$ or $z_\gamma$), the three approximations discussed in the previous section agree nicely. This is due to the fact that only a small fraction of events in these kinematics come from the near-encounter configurations where the geometrical effects become significant. As the photon energy is increased, so does the fraction of events from the near-encounter configurations, and eventually the predictions from the three predictions begin to diverge. In the bin of highest $z_\gamma$ this results in a 20\% enhancement in the WS compared to the PL approximation and, quite strikingly, a factor of two difference between the WS and WS$_{\delta({\bf s})}$ approximations. The latter clearly indicates that taking into account the spatial extent of the target nucleus is truly needed for an accurate interpretation of this observable.

\section{Modelling the e.m.\ breakup for the neutron-class selection}

\begin{figure}
  \centering
  \raisebox{-\height}{\includegraphics[width=0.495\textwidth]{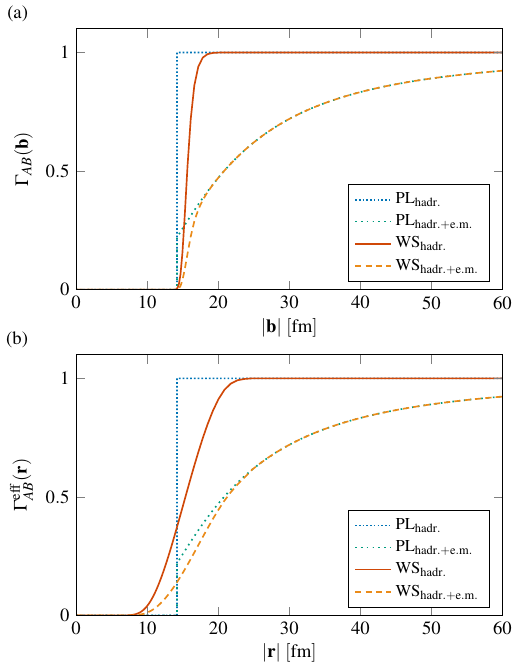}}
  \raisebox{-\height}{\includegraphics[width=0.495\textwidth]{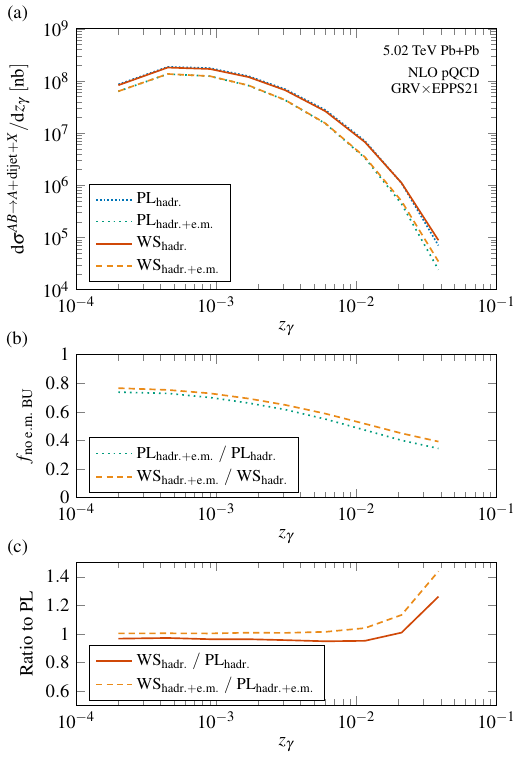}}
  \caption{Left: (a) The survival factor in pointlike and Wood-Saxon approximations, without and with the e.m.\ breakup factor. (b) Same, but for the effective $|{\bf r}|$-dependent survival factor. Right: The impact of the e.m.\ breakup correction on the dijet production. (a) Absolute cross section. (b) The size of the e.m.\ breakup correction. (c) The ratio between the predictions in pointlike and Wood-Saxon approximations, without and with the e.m.\ breakup factor. Figures from Ref.~\protect\citelow{Eskola:2024fhf}.}
  \label{fig:xsec-w-bu}
\end{figure}

Our treatment of the UPC dijet production in the previous section was fully inclusive, apart from requiring no direct hadronic interaction between the incoming nuclei. The experimental measurement however uses a 0nXn forward-neutron event-class selection and associated rapidity cuts for isolating the photonuclear events from generic nucleus-nucleus collisions.\cite{ATLAS:2022cbd} To account for the requirement of zero neutrons in the photon-going direction, we employ the Poissonian probability for no electromagnetic breakup of nucleus $A$ through Coulomb excitations~\cite{Baltz:2002pp}
\begin{equation}
  \Gamma_{AB}^{\rm e.m.}({\bf b}) = \exp\left[ - \int_0^1 {\rm d}y f_{\gamma/B}(y,{\bf b}) \sigma_{\gamma A \rightarrow A^*}(\sqrt{y\,s_{\rm NN}}) \right]
  \label{eq:em_bu}
\end{equation}
which we take from the Starlight event generator.\cite{Klein:2016yzr} The total (hadr.+e.m.) survival probability then reads
\begin{equation}
  \Gamma_{AB}^{\rm hadr.+e.m.}({\bf b}) = \Gamma_{AB}^{\rm e.m.}({\bf b}) \Gamma_{AB}^{\rm hadr.}({\bf b}), \qquad \Gamma_{AB,{\rm eff}}^{\rm hadr.+e.m.}({\bf r}) = \frac{1}{B} \int {\rm d}^2{\bf s} \, T_{B}({\bf s}) \, \Gamma_{AB}^{\rm hadr.+e.m.}({\bf r}\!-\!{\bf s}),
\end{equation}
where the latter form is the effective suppression factor multiplying the bare flux in Eq.~\eqref{eq:eff_flux_r_dep}. These are presented in Fig.~\ref{fig:xsec-w-bu} left, panels (a) and (b), respectively, without and with the e.m.\ breakup factor, for the PL and WS approximations. Requiring no e.m.\ breakup causes a clear, impact-parameter dependent reduction in the survival probability.

The impact on the dijet production is shown in Fig.~\ref{fig:xsec-w-bu} right, panels (a) and (b), where we see that taking into account the e.m.\ breakup yields a substantial suppression in the cross section. The impact-parameter dependence manifests itself here as a dependence in $z_\gamma$, and the suppression varies from 20\% at the lowest to 60\% at the highest $z_\gamma$. Noteworthily, this additional suppression does not reduce the difference between the PL and WS approximations of the effective flux, but rather enhances it and results in a 40\% difference in the highest $z_\gamma$ bin, as shown in Fig.~\ref{fig:xsec-w-bu} right, panel (c). Note that in order to match with the $X \geq 1$ neutron condition in the target-going direction, we should still subtract a diffractive contribution from our inclusive results, but we expect this to be a small correction in most of the phase-space.\cite{Eskola:2024fhf,Guzey:2020ehb}

\section{Spatial dependence in the nuclear modifications}

\begin{figure}
  \centering
  \raisebox{-\height}{\includegraphics[width=0.495\textwidth]{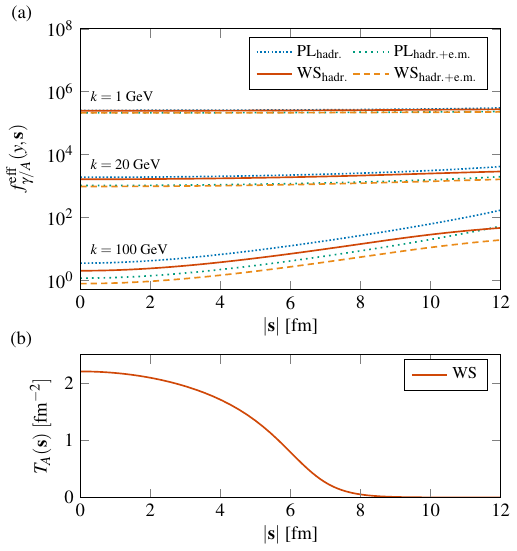}}
  \raisebox{-\height}{\includegraphics[width=0.495\textwidth]{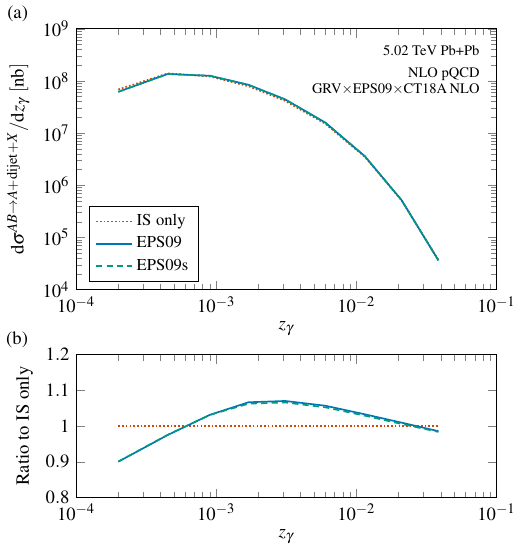}}
  \caption{Left: (a) The ${\bf s}$-dependent effective flux. (b) The nuclear thickness function. Right: (a) The UPC dijet cross section with predictions using either no nuclear modifications (IS only), spatially independent nuclear modifications (EPS09) or spatially dependent nuclear modifications (EPS09s). (b) Ratios to the IS only prediction. Figures from Ref.~\protect\citelow{Eskola:2024fhf}.}
  \label{fig:xsec-w-EPS09s}
\end{figure}

In above, we have treated the nPDFs under the factorization assumption $f_{j/B}(x,Q^2,{\bf s}) = \frac{1}{B} \, T_{B}({\bf s}) \times f_{j/B}(x,Q^2)$. This is of course just a simplification, and one should use proper impact-parameter dependent nPDFs instead.\cite{Frankfurt:2011cs,Helenius:2012wd} To study this, we begin by rewriting Eq.~\eqref{eq:xsec_full} as
\begin{equation}
  {\rm d} \sigma^{AB \rightarrow A + {\rm dijet} + X} \!=\! \sum_{i,j,X'} \int {\rm d}^2{\bf s} \, f_{\gamma/A}^{\rm eff}(y,{\bf s}) \otimes f_{i/\gamma}(x_\gamma,Q^2) \otimes f_{j/B}(x,Q^2,{\bf s})
  \otimes {\rm d} \hat{\sigma}^{ij \rightarrow {\rm dijet} + X'}(x_\gamma y p_A,x p_B,Q^2),
  \label{eq:xsec_re}
\end{equation}
thus making no assumption on the form of $f_{j/B}(x,Q^2,{\bf s})$, and where
\begin{equation}
  f_{\gamma/A}^{\rm eff}(y,{\bf s}) = \int {\rm d}^2{\bf r} \, \Gamma_{AB}({\bf r}\!-\!{\bf s}) \, f_{\gamma/A}(y,{\bf r})
  \label{eq:eff_flux_s_dep}
\end{equation}
is an ${\bf s}$-dependent effective photon flux. Should the spatial dependence integrate out, as happens in the far-passing region, this function would be constant in $|{\bf s}|$, whereas any sensitivity to the spatial dependence of the target nPDF would appear as a non-constant behaviour. This function is shown in Fig.~\ref{fig:xsec-w-EPS09s} left, panel (a), and compared with the target thickness function shown in Fig.~\ref{fig:xsec-w-EPS09s} left, panel (b). We see that, as could be expected from the discussion above, the ${\bf s}$-dependent effective flux is almost constant for low-energy photons where most of the contribution come from far-passing events, but for high-energy photons an enhancement towards the edge of the target appears due to an increasing contribution from the near-encounter region.

To test whether this $|{\bf s}|$-dependence of the flux at high photon energies can be used to probe the spatial dependence of nuclear modification, we employ the phenomenological EPS09s parametrization.\cite{Helenius:2012wd} The cross section can then be written as
\begin{multline}
  {\rm d} \sigma^{AB \rightarrow A + {\rm dijet} + X} = \sum_{i,j,X'} \sum_{m=0}^4 f_{\gamma/A}^{{\rm eff},m}(y) \otimes f_{i/\gamma}(x_\gamma,Q^2) \otimes f_{j/B}^m(x,Q^2)
  \otimes {\rm d} \hat{\sigma}^{ij \rightarrow {\rm dijet} + X'}(x_\gamma y p_A,x p_B,Q^2),
  \label{eq:xsec_EPS09s}
\end{multline}
where
\begin{equation}
  f_{\gamma/A}^{{\rm eff},m}(y) = \frac{1}{B} \int {\rm d}^2{\bf s} \, f_{\gamma/A}^{\rm eff}(y,{\bf s}) \, [T_{B}({\bf s})]^m, \qquad f_{j/B}^m(x,Q^2) = \sum_N c_m^{j/N}(x,Q^2) f_{j/N}(x,Q^2)
\end{equation}
are a generalised effective flux for different powers of the nuclear thickness function and the nuclear PDFs with EPS09s nuclear modification coefficients $c_m^{j/N}$ (where the sum goes over the nucleons $N$ in the nucleus $B$), respectively. The resulting dijet cross section is shown in Fig.~\ref{fig:xsec-w-EPS09s} right, panels (a) and (b), comparing
the prediction with spatially dependent nuclear modifications (EPS09s) to a version of the same nPDFs with no spatial dependence in the nuclear modifications (EPS09)\,\cite{Eskola:2009uj} and a prediction with no nuclear modifications in the PDFs but taking into account the trivial isospin dependence (IS only). As we can see, the nuclear modifications result in an order of 10\% effect in the cross section (EPS09 vs.\ IS only), while the spatial dependence in them gives only a small correction to that (EPS09s vs.\ EPS09). Hence, even though the spatial resolution at high $z_\gamma$ caused by the dominance of near-encounter events was able to distinguish the overall shape of the target nucleus (cf.\ the large difference between WS and WS$_{\delta({\bf s})}$ predictions), it is not strong enough to probe the spatial dependence of the nuclear modifications in impact-parameter dependent nPDFs.

\section{Conclusions}

In summary, the inclusive dijet photoproduction in UPCs provides valuable insights into nPDFs, but a careful consideration of the collision geometry and photon flux is essential for robust predictions and meaningful constraints. Even though we found that the spatial dependence in the nuclear modifications as encoded in the EPS09s nPDFs gave only a very small correction compared to the spatially independent EPS09 nuclear modifications, one should note that there is still a significant 40\% correction at large $z_\gamma$ from using a full WS calculation for the effective photon flux compared to the simple PL approximation, in the case when the e.m.\ breakup probability is properly taken into account. This e.m.\ survival factor was needed in order to match with the experimental 0nXn event selection.\cite{ATLAS:2022cbd} We also note that further measurements in the XnXn and 0n0n event classes will be helpful for experimentally quantifying the size of this e.m.\ survival factor and the diffractive contribution, and guide the reader to Ref.~\citelow{Eskola:2024fhf} for further discussion.

\section*{Acknowledgments}

This research was funded through the Research Council of Finland project No. 330448, as a part of the Center of Excellence in Quark Matter of the Research Council of Finland (projects No. 346325 and No. 346326) and as a part of the European Research Council project ERC-2018-ADG-835105 YoctoLHC. We acknowledge computing resources from the Finnish IT Center for Science (CSC), utilised under the project jyy2580.

\end{document}